\documentclass[twocolumn,showpacs,amsmath,amssymb]{revtex4}
\usepackage{amssymb}
\usepackage[dvips]{graphicx}
\begin{document}

\title{Combination quantum oscillations
in  canonical  single-band Fermi liquids}
\author{A. S. Alexandrov$^1$ and   V. V. Kabanov$^{2}$}

\affiliation{$^1$Department of Physics, Loughborough University,
Loughborough, United Kingdom\\ $^{2}$Josef Stefan Institute 1001,
Ljubljana, Slovenia}

\begin{abstract}
 Chemical potential oscillations  mix  individual-band frequencies
  of  the de Haas-van Alphen (dHvA) and Shubnikov-de
Haas (SdH) magneto-oscillations in canonical low-dimensional
\emph{multi}-band Fermi liquids. We predict a similar mixing  in
canonical \emph{single}-band Fermi liquids, which  Fermi-surfaces
have two or more extremal cross-sections. Combination harmonics are
analysed using a single-band almost two-dimensional energy spectrum.
We outline some experimental conditions allowing for resolution of
combination harmonics.
\end{abstract}

\maketitle

Magnetic quantum oscillations of  magnetisation (dHvA effect) and
resistivity (SdH effect) are  unequivocal hallmarks of the
Fermi-liquid, providing  most reliable and detailed Fermi-surfaces
\cite{shon}, in particular in layered
 organic metals \cite{sin,kar} and almost two-dimensional (2D)  superconductors
 like Sr$_2$RuO$_4$ \cite{mac}. An interesting feature
 of dHvA/SdH oscillations  is a measurable difference between canonical and
grand canonical ensembles, which is most pronounced  in multi-band
 low-dimensional metals \cite{alebra}. Thermodynamically two
ensembles must be identical, but quantum fluctuations are
fundamentally different depending on whether  measurements are
performed on either closed or open system with fixed electron
density, $n_e$,  or chemical potential, $\mu$, respectively. The
difference between  corresponding free energies is tiny, since it is
proportional to fluctuations of the carrier density,  however, the
effect on quantum corrections in magnetisation and  conductivity is
significant.

 In particular, there are  combination frequencies in
dHvA/SdH oscillations of a two-dimensional multiband metal with
fixed $n_e$,
 predicted   by Alexandrov and Bratkovsky (AB) \cite {alebra}, and
 studied  numerically \cite{alebra,nak,alebra2,cham,comment,taut} and
analytically \cite{alebra3,alebra4,fort}.   The effect was
experimentally observed in different low-dimensional systems
\cite{she,other,kar}. Obviously, there are no chemical potential
oscillations when $\mu$ is fixed by a reservoir, so there is no
mixing of the individual-band  fundamental frequencies in the
Fourier transform (FT) of magnetisation   in an open
(grand-canonical) system. Importantly, samples are normally placed
on non-conducting substrates with no electrodes attached, so the
system is closed in actual dHvA experiments.

As it happens  the fundamental frequency mixing due to the chemical
potential oscillations (AB effect) may be obscured by  mixing due to
the magnetic breakdown \cite{MB} (MB effect), as discussed by
Kartsovnik \cite{kar}. The MB effect is the switching of two close
electron orbits in different bands on the Fermi-surface (FS)  at
sufficiently strong magnetic fields. Here we predict a mixing of two
or more fundamental frequencies in a canonical \emph{single}-band
Fermi liquid with a few extremal FS cross-sections, where the MB is
non-existent.

To illustrate the point we consider an anisotropic single band, with
the dispersion, $E({\bf k})$ , in zero magnetic field,
\begin{equation}
E({\bf k})= {\hbar^2 k_\parallel^2\over{2m}}-2t_\perp\cos (k_\perp
d), \label{spec}
\end{equation}
 which is a fair approximation for a band in layered metals \cite{sin,kar}. Here
$k_\parallel$ and $k_\perp$ are the in-plane and out-of-plane
quasi-momenta, $t$ is the inter-plane hopping integral, and $d$ is
the inter-plane distance.

\begin{figure}
\begin{center}
\includegraphics[angle=-90,width=0.45\textwidth]{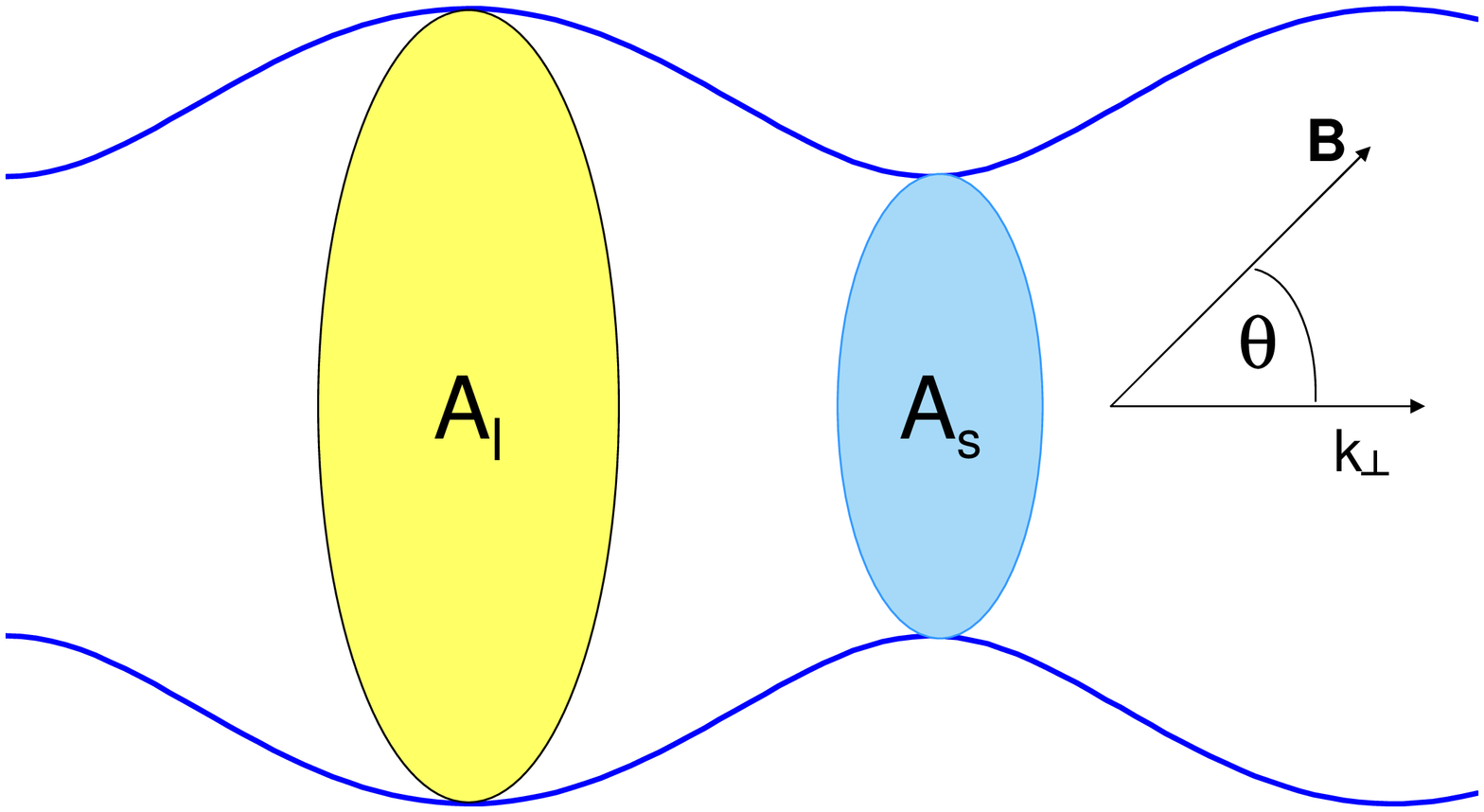}
\vskip -0.5mm \caption{ Large, $A_l=2\pi m(\mu+2t)/\hbar^2$, and
small, $A_s=2\pi m(\mu-2t)/\hbar^2$, extremal  cross-sections  of a
 layered-metal Fermi-surface. }
\end{center}
\end{figure}

When the magnetic field, $B$, is applied, the spectrum
Eq.(\ref{spec}) is quantised as \cite{kurihara}
\begin{equation}
E_n(k_\perp)= \hbar\omega_c(n+1/2) -2t\cos (k_\perp d)\pm g\mu_BB/2,
\label{spec2}
\end{equation}
where $\omega_c=eB\cos(\Theta)/m$ is the cyclotron frequency
($n=0,1,2,...$), $t=t_\perp J_0(k_Fd\tan(\Theta))$ ($J_0(x)$ is the
Bessel function),  $\Theta$ is the angle between the field and the
normal to the planes, $g$  is the electron g-factor, and $\mu_B$ is
the Bohr  magneton.  The spectrum, Eq.(\ref{spec2}) is perfectly 2D
at the Yamaji angles \cite{yam} found from
$J_0(k_Fd\tan(\Theta))=0$, where $\hbar k_F=(2m \mu)^{1/2}$ is the
Fermi momentum in pure 2D case, but otherwise there are two extremal
semiclassical orbits. They give rise to beats in dHvA/SdH
oscillations with two fundamental FT frequencies, $F_{l,s}=\hbar
A_{l,s}/2\pi e \cos(\Theta)$ , revealing modulations of the
cylindrical FS along the perpendicular direction, Fig.1, as observed
e.g. in Sr$_2$RuO$_4$ \cite{mac,other}.

Since there are no different bands one might  expect neither AB nor
MB mixing of the fundamental frequencies, $F_l$ and $F_s$  in the
single-band model, Eq.(\ref{spec}), in contrast with canonical
multi-band systems \cite{alebra,MB}. Actually, as we show below,
$F_l$ and $F_s$ turn out mixed, if $n_e$ is constant, so that a
combination  frequency $F_+=F_l+F_s$ appears  similar to the AB
combination frequency \cite{alebra} in two-band canonical
Fermi-liquids. Using  conventional Poisson's summation and integrals
\cite{shon}  the grand canonical potential per unit volume,
\begin{equation}
 \Omega=-{k_BT eB\cos(\Theta)\over{4\pi^2\hbar}}\sum_{n}
\int_{-\pi/d}^{\pi/d} dk_\perp \ln
[1+e^{(\mu-E_{n}(k_\perp))/k_BT}],
\end{equation}
is given by $\Omega=\tilde{\Omega}- m\mu^2/2\pi d\hbar^2,$ where
\begin{eqnarray}
\tilde{\Omega}&=&{e^2B^2\cos^2(\Theta)\over{4\pi^3 md
}}\sum_{r=1}^{\infty}R_T \left({2\pi^2 r k_BT \over{\hbar
\omega_c}}\right)\cos\left({\pi r
gm\over{m_e\cos(\Theta)}}\right)\cr &\times&
{(-R)^{r}\over{r^{2}}}J_0\left({4\pi r t \over{\hbar
\omega_c}}\right )\cos \left( \frac{2\pi r \mu}{\hbar
\omega_c}\right) \label{Omega}
\end{eqnarray}
is its quantum part with the conventional  temperature,
$R_T(x)=x/\sinh(x)$, and Dingle (i.e. collision),  $0< R\leqslant
1$, damping factors, as derived  in Ref. \cite{alebra4}.
Differentiating $\Omega$ with respect to the magnetic field at
constant $\mu$, one obtains the oscillating part of the
magnetisation, $\tilde{M}=-
\partial \tilde{\Omega}/\partial B$,
\begin{equation}
\tilde{M}={e\mu \cos(\Theta) \over{2\pi^2 \hbar
d}}\sum_{r=1}^{\infty}{(-R)^{r}\over{r}}J_0\left({4\pi  r t
\over{\hbar \omega_c}}\right )\sin \left( \frac{2\pi r \mu}{\hbar
\omega_c}\right), \label{M}
\end{equation}
where we neglect small terms of the order of $2t/\mu\ll 1$, and take
zero-temperature limit and $g=0$ for more transparency.

We are interested  in the regime $\hbar\omega_c \ll 4\pi t$, where
three-dimensional corrections to the spectrum are significant,
rather than in the opposite ultra-quantum limit \cite{alebra4},
where the quantised spectrum is almost 2D. In our intermediate-field
regime one can replace the Bessel function in Eq.(\ref{M}) by its
asymptotic, $J_0(x)\approx (2/\pi x)^{1/2} \cos (x-\pi/4)$ at large
$x$ to obtain
\begin{eqnarray}
\tilde{M}&=&{e\mu \cos(\Theta) \over{4\pi^2 \hbar
d}}\left({2B\over{\pi B_\perp}}\right)^{1/2}
\sum_{r=1}^{\infty}{(-R)^{r}\over{r^{3/2}}}\cr &\times& [\sin \left(
\frac{2\pi r F_l}{B}-{\pi \over{4}}\right)+\sin \left( \frac{2\pi r
F_s}{B}+ {\pi \over{4}}\right )], \label{MM}
\end{eqnarray}
where $B_{\perp}\equiv 4\pi mt/e\hbar \cos(\Theta) \gg B$.

Naturally the FT of Eq.(\ref{MM}) yields two fundamental frequencies
in the grand-canonical ensemble, where $\mu$ is fixed, Fig.2.
\begin{figure}
\begin{center}
\includegraphics[angle=-0,width=0.53\textwidth]{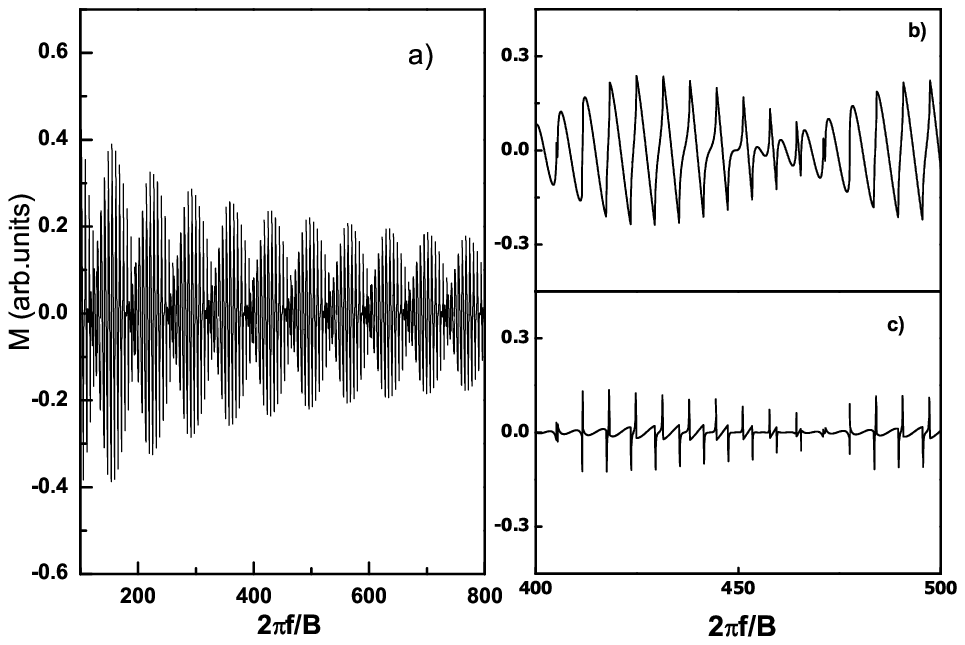}
\includegraphics[angle=-0,width=0.53\textwidth]{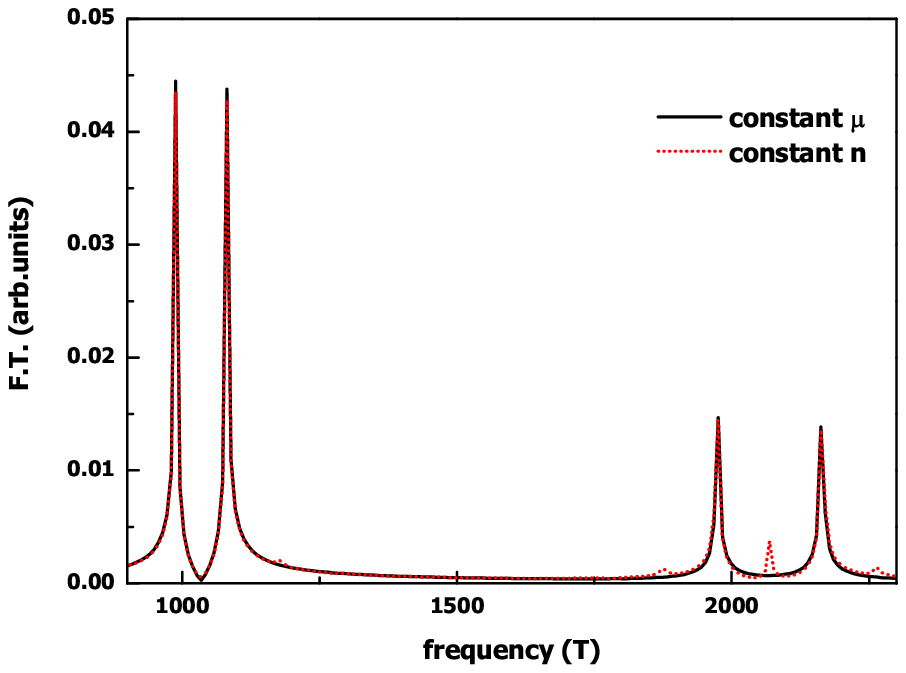}
\vskip -0.5mm \caption{Upper panel: magnetisations as  functions of
 the inverse magnetic field $1/B$ for  grand-canonical (a) and  canonical
(b) single-band Fermi-liquids, and their difference (c); lower
panel: magnetisation FTs for grand-canonical (solid line) and
canonical (dashed line) single-band Fermi-liquids. Here $2\pi f=\pi
(f_l+f_s) = 6500$ Tesla, $B_\perp=300$ Tesla, $R=1$ and $T=g=0$. }
\end{center}
\end{figure}
However, the chemical potential  oscillates with the magnetic field
in the canonical system \cite{shon,alebra}, which affects quantum
corrections to magnetisation.  Using $n_e= -\partial \Omega/\partial
\mu$, one  can find the oscillating component, $\tilde{\mu}\equiv z
\omega_c/2\pi=\partial \tilde{\Omega}/\partial \mu$, of the chemical
potential,  $\mu=\mu_0+\tilde{\mu}$, where $\mu_0= d\pi \hbar^2
n_e/m$ is its zero-field value and
\begin{eqnarray}
z&=&\left({B\over{2\pi B_\perp}}\right)^{1/2}
\sum_{r=1}^{\infty}{(-R)^{r}\over{r^{3/2}}}\times  \label{MMM} [\sin
\left(r z + \frac{2\pi r f_l}{B}-{\pi \over{4}}\right) \cr &+&\sin
\left(r z+ \frac{2\pi r f_s}{B}+ {\pi \over{4}}\right )].\label{MMM}
\end{eqnarray}
Here  the "bare" fundamental frequencies, $f_{l,s}= m(\mu_0 \pm
2t)/e\hbar \cos(\Theta)$, are now field-independent. Remarkably,
apart from a normalising factor, the dimensionless quantum
correction, $z$, to the chemical potential, Eq.(\ref{MMM}), turns
out identical to the magnetisation quantum correction, $\tilde{M}$,
Eq.(\ref{MM}), which is not the case in a two-band canonical Fermi
liquid \cite{comment}.

To get insight regarding the  FT of $z(B)$ or $\tilde{M}(B)$,
Eq.(\ref{MMM}), we first apply an   analytical perturbative approach
of Refs. \cite{alebra3,comment} expanding $z$ in powers of $R$ up to
the second order, $z \approx z_1+z_2+z_{mix}$, where
\begin{eqnarray}
z_1&=&-R\left({B\over{2\pi B_\perp}}\right)^{1/2}
 [\sin \left(\frac{2\pi r
f_l}{B}-{\pi \over{4}}\right)\cr &+&\sin \left( \frac{2\pi r
f_s}{B}+ {\pi \over{4}}\right )]\label{M1}
\end{eqnarray}
yields  two first fundamental harmonics with the frequencies $f_l$
and $f_s$ identical to those of the grand-canonical system,
\begin{eqnarray}
z_2 &\approx& {R^2\over{2^{3/2}}}\left({B\over{2\pi
B_\perp}}\right)^{1/2}
 [\sin \left(\frac{4\pi r
f_l}{B}-{\pi \over{4}}\right) \cr &+&\sin \left( \frac{4\pi r
f_s}{B}+ {\pi \over{4}}\right )]\label{M2}
\end{eqnarray}
yields two  second fundamental harmonics with the frequencies $2f_l$
and $2f_s$ as in the grand-canonical system, and
\begin{equation}
z_{mix}= R^2{B\over{2\pi B_\perp}}
 \sin \left(\frac{2\pi
F_+}{B}\right)\label{M3}
\end{equation}
is the mixed harmonic with the frequency $F_+= f_l+f_s$, which is a
specific signature of the canonical ensemble. Its amplitude is small
compared with the first-harmonics amplitudes  as $ R(B/2\pi
B_{\perp})^{1/2}$ in contrast with  multi-band systems, where the
mixed-harmonic amplitudes have roughly  the same order of magnitude
as the fundamental-harmonic amplitudes (at $R=1$)
\cite{alebra,alebra3}. Also there is no $F_-=f_l-f_s$ frequency in
the FT spectrum of the single-band canonical system, different from
the multi-band canonical systems \cite{nak,alebra2}.

To assess an accuracy of the analytical approximation,
Eq.(\ref{M3}), and some experimental conditions, allowing for
resolution of the mixed harmonic, we present numerically exact
magnetisation and their FTs in Fig.2. Since convergence of the sum
in Eq.(\ref{MMM}) is poor, one can use its integral representation
in numerical calculations as
\begin{eqnarray}
z&=&\left({B\over{2\pi B_\perp}}\right)^{1/2} \Im [e^{-i\pi /4}
Li_{3/2}(-Re^{i(z+2\pi f_l/B)})\cr &+& e^{i\pi
/4}Li_{3/2}(-Re^{i(z+2\pi f_s/B)})],\label{MMMM}
\end{eqnarray}
where $Li_{3/2}(x)=(2/\pi^{1/2}) \int_0^{\infty} dt t^{1/2}/(e^{
t}/x-1)$ is the polylogarithm. The analytical amplitudes,
Eqs.(\ref{M1},\ref{M2},\ref{M3}), prove to be practically exact with
the relative error below 10 percent at any $R$, Fig.3,  as the
amplitudes of the analytical theory of dHvA effect in canonical
multi-band systems \cite{alebra3,comment}.
\begin{figure}
\begin{center}
\includegraphics[angle=0,width=0.50\textwidth]{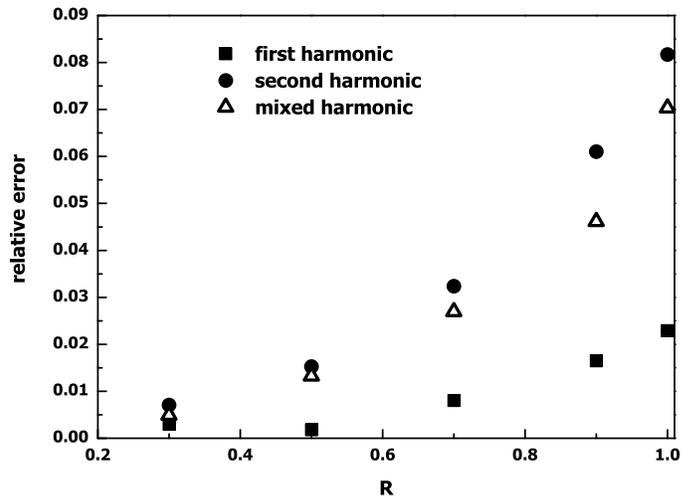}
\vskip -0.5mm \caption{ Relative errors of analytical harmonic
amplitudes, Eqs.(\ref{M1}, \ref{M2}, \ref{M3})  with respect to
numerically exact amplitudes. }
\end{center}
\end{figure}
Another important feature of the numerical FT of the solution of
Eq.(\ref{MMMM}) is that the resolution of the mixed central peak
 in the middle between two fundamental second  harmonics, Fig.2
(lower panel), essentially depends on the magnetic-field window used
in FT, Fig.4.
\begin{figure}
\begin{center}
\includegraphics[angle=0,width=0.50\textwidth]{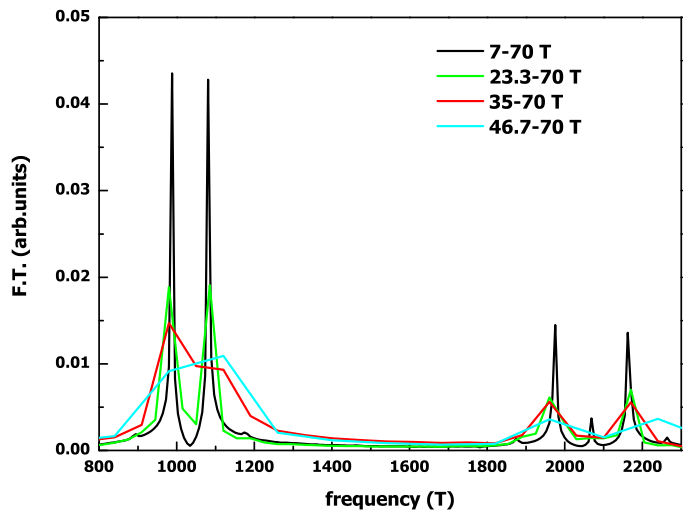}
\vskip -0.5mm \caption{ Effect of the magnetic field window on the
Fourier transform of magnetisation. Decreasing the window increases
the width of  FT harmonics obscuring the mixed harmonic.  }
\end{center}
\end{figure}
Since the mixed amplitude is relatively small as $(B/2\pi
B_\perp)^{1/2} \ll 1$, the window affects  its experimental
resolution. We believe that a relatively small interval of the
magnetic fields, used in FT, has  prevented so far the
\emph{single-band} combination frequency to be seen in layered
metals \cite{sin,kar,mac,other}. Importantly, since the
characteristic field, $B_\perp \propto
J_0(k_Fd\tan(\Theta))/\cos(\Theta)$, is an oscillating function of
the tilting angle $\Theta$, the combination amplitude  also
oscillates as the function of the angle, which could be instrumental
in its experimental identification. We notice that the angle
dependence of the second fundamental harmonics has been  clearly
observed in Sr$_2$RuO$_4$ \cite{other}.

There is also mixing  in the SdH quantum oscillations of transverse
and longitudinal conductivities. For example the longitudinal
conductivity is given by the Kubo formula \cite{kubo},
\begin{equation}
\sigma =-\pi \hbar e^2 \int dE {\partial f (E)\over{\partial E}} Tr
[ \delta (E-H)v_\perp\delta (E-H)v_\perp)], \label{Kubo}
\end{equation}
where $v_\perp =2t d \sin(k_\perp d)/\hbar$ is the longitudinal
component of the velocity operator,  $H$ is the  single-particle
Hamiltonian including the impurity scattering, and
$f(E)=1/[\exp(E-\mu)/k_BT+1]$ is the Fermi-Dirac distribution
function. Averaging over random impurity distributions and
approximating the scattering rate by a constant, $\Gamma$, one
obtains the trace in Eq.(\ref{Kubo}) as $(v_\perp \Im
[E-E_n(k_\perp)-i\Gamma]^{-1})^2$ in the ladder approximation. Then
applying  Poisson's summation, one can readily obtain a quantum
correction, $\tilde{\sigma}$, to the classical conductivity (for
 detail see  Refs. \cite{min,gri,alekab}), which is (at $T=0$)
\begin{equation}
\tilde{\sigma} \propto B
\sum_{r=1}^{\infty}{(-R)^{r}\over{r}}J_1\left({4\pi r t \over{\hbar
\omega_c}}\right )\cos \left( \frac{2\pi r \mu}{\hbar
\omega_c}\right). \label{M}
\end{equation}
The asymptotic of the Bessel function, $J_1(x)\approx (2/\pi
x)^{1/2} \cos (x+\pi/4)$, yields FTs of $\tilde{\sigma}$ very
similar to those of magnetisation, Fig.2, with the combination
harmonic in the canonical system. Generally the scattering rate
depends on the magnetic field \cite{gri}, so that its oscillations
require more thorough analysis of the SdH effect, but mixing should
be robust. Interestingly, some mixing of fundamental frequencies may
occur even in  grand-canonical
 multi- or single-band layered systems, if there is an inter-band
 or inter-extremal cross-section scattering  by impurities.

In conclusion, we have found the combination frequency in the
quantum magnetic oscillations of the single-band canonical layered
Fermi liquid. The difference between quantum oscillations of the
canonical and grand-canonical ensembles,  is tiny, Fig.2,  but not
obscured by the MB effect, which is absent in the single-band case
in  contrast with   the multi-band systems.    We have also shown
that the analytical (perturbative) FT amplitudes  are numerically
accurate even at zero temperature and in  clean samples (i.e. for
$R=1$) as they are in the multi-band analytical theory
\cite{alebra3,comment}. A wide magnetic-field window is essential
for experimental resolution of the combination dHvA/SdH frequency.

We greatly appreciate valuable discussions with Iorwerth Thomas and
support of this work by EPSRC (UK) (grant No. EP/D035589).


\begin{thebibliography}{90}
\bibitem{shon}
D. Schoenberg, \emph{Magnetic Oscillations in Metals} (Cambridge
University Press, Cambridge 1984).
\bibitem{sin}
J. Singleton, Rep. Prog. Phys. {\bf 63}, 1111 (2000).
\bibitem{kar}
M. V. Kartsovnik, Chem. Rev. {\bf 104}, 5737 (2004) and references
therein.
\bibitem{mac} A. P. Mackenzie, S. R. Julian, A. J. Diver, G. J. McMullan, M. P. Ray,
G. G. Lonzarich, Y. Maeno, S. Nishizaki, and T. Fujita, Phys. Rev. Lett. {\bf 76}, 3786
(1996).
\bibitem{alebra}  A. S. Alexandrov and A. M. Bratkovsky, Phys. Rev. Lett. {\bf %
76, }1308 (1996).
\bibitem{nak}  M. Nakano, J. Phys. Soc. Japan {\bf 66,} 19 (1997).
\bibitem{alebra2}  A. S. Alexandrov and A. M. Bratkovsky, Phys. Lett. A {\bf %
234,} 53 (1997).
\bibitem{cham}  T. Champel, Phys. Rev. B {\bf 65, }153403 (2002);
ibid {\bf 69}, 167402 (2004).
\bibitem{comment} A. S. Alexandrov and A. M. Bratkovsky, Phys. Rev. B {\bf %
69, }167401 (2004).
\bibitem{taut} K. Kishigi and Y. Hasegawa, Phys.  Rev. B {\bf
65}, 205405 (2002); ibid. {\bf 72}, 045410 (2005).
\bibitem{alebra3}  A. S. Alexandrov and A. M. Bratkovsky, Phys. Rev. B {\bf %
63, }033105 (2001).
\bibitem{alebra4}  A. M. Bratkovsky and A. S. Alexandrov, Phys. Rev. B {\bf %
65, }035418 (2002).
\bibitem{fort} J. Y. Fortin, E. Perez,
and A. Audouard,  Phys. Rev. B {\bf 71}, 155101 (2005).
\bibitem{she}  R. A. Shepherd, M. Elliott, W. G. Herrenden-Harker, M. Zervos,
P. R. Morris, M. Beck, and M. Ilegems, Phys. Rev. B {\bf 60,} R11277
(1999).
\bibitem{other}  E. Ohmichi, Y. Maeno, and T. Ishiguro, J. Phys. Soc Japan {\bf  68},
24 (1999).
\bibitem{MB}
N. Harrison, J. Caulfield, J. Singleton, P. H. P. Reinders, F.
Herlach, W. Hayes, M. Kurmoo, and P. J. Day, J. Phys. Condens.
Matter {\bf 8}, 5415 (1996); P. S. Sandhu, Ju H. Kim, and J. S.
Brooks, Phys. Rev. B {\bf 56}, 11566 (1997);
 J. H. Kim, S. Y. Han, and J. S. Brooks,
Phys. Rev. B {\bf 60}, 3213 (1999); S. Y. Han, J. S. Brooks, and J.
H. Kim, Phys. Rev. Lett. {\bf 85}, 1500 (2000); V. M. Gvozdikov and
M. Taut, Phys. Rev. B {\bf 75}, 155436 (2007); D. Vignolles, A.
Audouard, V. N. Laukhin, J. Beard, E. Canadell, N. G. Spitsina, E.
B. Yagubskii, Eur. Phys. J. B {\bf 55}, 383 (2007).
\bibitem{kurihara} Y. Kurihara, J. Phys. Soc. Japan {\bf  61},
975 (1989).

\bibitem{yam} K. Yamaji, J. Phys. Soc. Japan {\bf  58},
1520 (1989).
\bibitem{kubo}
 R. Kubo, H. Hasegava, and N. Hashitsume,
J. Phys. Soc. Japan {\bf 14}, 56 (1959)).

\bibitem{min} T. Champel and V. P. Mineev, Phys. Rev. B {\bf 66},
195111 (2002); ibid {\bf 67}, 089901 (2003).
\bibitem{gri} P. D. Grigoriev, Phys. Rev. B {\bf 67}, 144401 (2003).
\bibitem{alekab}
A. S. Alexandrov and V. V. Kabanov, Phys. Rev. Lett. {\bf 95},
076601 (2005); ibid, 169902 (2005).



\end{thebibliography}
\end{document}